\documentclass[floatfix,secnumarabic,amssymb,bibnotes, aps,prl,reprint,amsmath,superscriptaddress]{revtex4-2}
\usepackage{epsfig}
\usepackage{graphicx}% Include figure files
\usepackage{dcolumn}% Align table columns on decimal point
\usepackage{bm}% bold math
\usepackage{CJK}
\usepackage{float}
\usepackage{tabularx}
\usepackage{threeparttable}
\usepackage{multirow}
\usepackage{xcolor}
\begin{document}
\begin{CJK}{GBK}{song}
\title{Quasi-free Neutron Knockout Reaction Reveals a Small $s$-orbital Component in the Borromean Nucleus $^{17}$B}
\def \cns      {Center for Nuclear Study, University of Tokyo, Hongo 7-3-1, Bunkyo, Tokyo 113-0033, Japan}
\def \rnc      {RIKEN Nishina Center, Hirosawa 2-1, Wako, Saitama 351-0198, Japan}
\def \saclay   {D\'{e}partement de Physique Nucl\'{e}aire, IRFU, CEA, Universit\'{e} Paris-Saclay, F-91191 Gif-sur-Yvette, France}
\def \tud      {Institut f\"{u}r Kernphysik, Technische Universit\"{a}t Darmstadt, D-64289 Darmstadt, Germany}
\def \peking   {School of Physics and State Key Laboratory of Nuclear Physics and Technology, Peking University, Beijing 100871, China}
\def \orsay    {IPN Orsay, Universit\'{e} Paris Sud, IN2P3-CNRS, F-91406 Orsay Cedex, France}
\def \caen     {LPC Caen, ENSICAEN, Universit\'{e} de Caen Normandie, CNRS/IN2P3, F-14050 Caen Cedex, France}
\def \tohoku   {Department of Physics, Tohoku University, Aramaki Aza-Aoba 6-3, Aoba, Sendai, Miyagi 980-8578, Japan}
\def \miyazaki {Department of Applied Physics, University of Miyazaki, Gakuen-Kibanadai-Nishi 1-1, Miyazaki 889-2192, Japan}
\def \ehwa     {Department of Physics, Ehwa Womans University, Seoul 03760, Republic of Korea}
\def \ut       {Department of Physics, University of Tokyo, Hongo 7-3-1, Bunkyo, Tokyo 113-0033, Japan}
\def \titech   {Department of Physics, Tokyo Institute of Technology, 2-12-1 O-Okayama, Meguro, Tokyo 152-8551, Japan}
\def \atomki   {Institute for Nuclear Research, Hungarian Academy of Sciences (MTA Atomki), P.O. Box 51, H-4001 Debrecen, Hungary}
\def \kyoto    {Department of Physics, Kyoto University, Kitashirakawa, Sakyo, Kyoto 606-8502, Japan}
\def \kyushu   {Department of Physics, Kyushu University, Nishi, Fukuoka 819-0395, Japan}
\def \tum      {Physik Department, Technische Universit\"{a}t M\"{u}nchen, D-85748 Garching, Germany}
\def \rcnp     {Research Center for Nuclear Physics (RCNP), Osaka University, 10-1 Mihogaoka, Ibaraki, Osaka 567-0047, Japan}
\def \ikp      {Institut f\"{u}r Kernphysik, Technische Universit\"{a}t Darmstadt, D-64289 Darmstadt, Germany}
\def \cenm     {Center for Extreme Nuclear Matters, Korea University, Seoul 02841, Republic of Korea}
\def \ocu      {Department of Physics, Osaka City University, Osaka 558-8585, Japan}
\def \nit      {Tokuyama College, National Institute of Technology, Yamaguchi 745-8585, Japan}
\def \york     {Department of Physics, University of York, Heslington, York YO10 5DD, United Kingdom}
\def \lbnl     {Nuclear Science Division, Lawrence Berkeley National Laboratory, Berkeley, California 94720, USA}
\def \cpr      {Cluster for Pioneering Research, RIKEN, Hirosawa 2-1, Wako, Saitama 351-0198, Japan}
\def \sys  {Sino-French Institute of Nuclear Engineering and Technology, Sun Yat-Sen University, Zhuhai, 519082, Guangdong, China} %C.X Yuan

\def \itp {CAS Key Laboratory of Theoretical Physics, Institute of Theoretical Physics, Chinese Academy of Sciences, Beijing 100190, China} %X.X Sun
\def \ucasa {School of Physical Sciences, University of Chinese Academy of Sciences, Beijing 100049, China} %X.X Sun
\def \imp {Institute of Modern Physics, Chinese Academy of Sciences, Lanzhou 730000, China}
\def \ucasb {School of Nuclear Science and Technology, University of Chinese Academy of Sciences, Beijing 100049, China} % Nicolas Michel,Wei Zuo
\def \jaea {Advanced Science Research Center, Japan Atomic Energy Agency, Tokai, Ibaraki 319-1195, Japan}

\def \hokku {Department of Physics, Hokkaido University, Sapporo 060-0810, Japan}
\def \hokkub {Nuclear Reaction Data Centre, Hokkaido University, Sapporo 060-0810, Japan}
\def \rikkyo {Department of Physics, Rikkyo University, 3-34-1, Nishi-Ikebukuro, Toshima, Tokyo 171-8501, Japan}
\def \ibs {Center for Exotic Nuclear Studies, Institute for Basic Science, Daejeon 34126, Republic of Korea}

\author{Z.~H.~Yang}\email{zhyang@ribf.riken.jp}\affiliation{\rcnp}\affiliation{\rnc}
\author{Y.~Kubota}\altaffiliation[Present address: ]{Institut f\"{u}r Kernphysik, Technische Universit\"{a}t Darmstadt, D-64289 Darmstadt, Germany}\affiliation{\rnc}\affiliation{\cns}
\author{A.~Corsi}\affiliation{\saclay}

\author{K.~Yoshida}\affiliation{\jaea}
\author{X.-X.~Sun}\affiliation{\itp}\affiliation{\ucasa}
\author{J.~G.~Li}\affiliation{\peking}
\author{M.~Kimura}\affiliation{\hokku}\affiliation{\hokkub}\affiliation{\rcnp}
\author{N.~Michel}\affiliation{\imp}\affiliation{\ucasb}
\author{K.~Ogata}\affiliation{\rcnp}\affiliation{\ocu}
\author{C.~X.~Yuan}\affiliation{\sys}
\author{Q.~Yuan}\affiliation{\peking}

\author{G.~Authelet}\affiliation{\saclay}
\author{H.~Baba}\affiliation{\rnc}
\author{C.~Caesar}\affiliation{\tud}
\author{D.~Calvet}\affiliation{\saclay}
\author{A.~Delbart}\affiliation{\saclay}
\author{M.~Dozono}\affiliation{\cns}
\author{J.~Feng}\affiliation{\peking}
\author{F.~Flavigny}\altaffiliation[Present address: ]{LPC Caen, ENSICAEN, Universit\'{e} de Caen Normandie, CNRS/IN2P3, F-14050 Caen Cedex, France}\affiliation{\orsay}
\author{J.-M.~Gheller}\affiliation{\saclay}
\author{J.~Gibelin}\affiliation{\caen}
\author{A.~Giganon}\affiliation{\saclay}
\author{A.~Gillibert}\affiliation{\saclay}
\author{K.~Hasegawa}\affiliation{\tohoku}
\author{T.~Isobe}\affiliation{\rnc}
\author{Y.~Kanaya}\affiliation{\miyazaki}
\author{S.~Kawakami}\affiliation{\miyazaki}
\author{D.~Kim}\affiliation{\ibs}
\author{Y.~Kiyokawa}\affiliation{\cns}
\author{M.~Kobayashi}\affiliation{\cns}
\author{N.~Kobayashi}\affiliation{\ut}
\author{T.~Kobayashi}\affiliation{\tohoku}
\author{Y.~Kondo}\affiliation{\titech}
\author{Z.~Korkulu}\affiliation{\ibs}\affiliation{\atomki}
\author{S.~Koyama}\affiliation{\ut}
\author{V.~Lapoux}\affiliation{\saclay}
\author{Y.~Maeda}\affiliation{\miyazaki}
\author{F.~M.~Marqu\'{e}s}\affiliation{\caen}
\author{T.~Motobayashi}\affiliation{\rnc}
\author{T.~Miyazaki}\affiliation{\ut}
\author{T.~Nakamura}\affiliation{\titech}
\author{N.~Nakatsuka}\affiliation{\kyoto}
\author{Y.~Nishio}\affiliation{\kyushu}
\author{A.~Obertelli}\altaffiliation[Present address: ]{Institut f\"{u}r Kernphysik, Technische Universit\"{a}t Darmstadt, D-64289 Darmstadt, Germany}\affiliation{\saclay}
\author{A.~Ohkura}\affiliation{\kyushu}
\author{N.~A.~Orr}\affiliation{\caen}
\author{S.~Ota}\affiliation{\cns}
\author{H.~Otsu}\affiliation{\rnc}
\author{T.~Ozaki}\affiliation{\titech}
\author{V.~Panin}\affiliation{\rnc}
\author{S.~Paschalis}\altaffiliation[Present address: ] {Department of Physics, University of York, Heslington, York YO10 5DD, United Kingdom}\affiliation{\tud}
\author{E.~C.~Pollacco}\affiliation{\saclay}
\author{S.~Reichert}\affiliation{\tum}
\author{J.-Y.~Rouss\'{e}}\affiliation{\saclay}
\author{A.~T.~Saito}\affiliation{\titech}
\author{S.~Sakaguchi}\affiliation{\kyushu}
\author{M.~Sako}\affiliation{\rnc}
\author{C.~Santamaria}\affiliation{\saclay}
\author{M.~Sasano}\affiliation{\rnc}
\author{H.~Sato}\affiliation{\rnc}
\author{M.~Shikata}\affiliation{\titech}
\author{Y.~Shimizu}\affiliation{\rnc}
\author{Y.~Shindo}\affiliation{\kyushu}
\author{L.~Stuhl}\affiliation{\ibs}\affiliation{\rnc}
\author{T.~Sumikama}\affiliation{\tohoku}
\author{Y.~L.~Sun}\altaffiliation[Present address: ]{Institut f\"{u}r Kernphysik, Technische Universit\"{a}t Darmstadt, D-64289 Darmstadt, Germany}\affiliation{\saclay}
\author{M.~Tabata}\affiliation{\kyushu}
\author{Y.~Togano}\affiliation{\titech} \affiliation{\rikkyo}
\author{J.~Tsubota}\affiliation{\titech}
\author{F.~R.~Xu}\affiliation{\peking}
\author{J.~Yasuda}\affiliation{\kyushu}
\author{K.~Yoneda}\affiliation{\rnc}
\author{J.~Zenihiro}\affiliation{\rnc}
\author{S.-G.~Zhou}\affiliation{\itp}\affiliation{\ucasa}
\author{W.~Zuo}\affiliation{\imp}\affiliation{\ucasb}
\author{T.~Uesaka}\affiliation{\rnc}\affiliation{\cpr}

%\date{\today}
%\date{}
\begin{abstract}
A kinematically complete quasi-free $(p,pn)$ experiment in inverse kinematics was performed to study the structure of the Borromean nucleus $^{17}$B, which had long been considered to have neutron halo. By analyzing the momentum distributions and exclusive cross sections, we obtained the spectroscopic factors for $1s_{1/2}$ and $0d_{5/2}$ orbitals, and a surprisingly small percentage of 9(2)$\%$ was determined for $1s_{1/2}$. Our finding of such a small $1s_{1/2}$ component and the halo features reported in prior experiments can be explained by the deformed relativistic Hartree-Bogoliubov theory in continuum, revealing a definite but not dominant neutron halo in $^{17}$B. The present work gives the smallest $s$- or $p$-orbital component among known nuclei exhibiting halo features, and implies that the dominant occupation of $s$ or $p$ orbitals is not a prerequisite for the occurrence of neutron halo.
\end{abstract}


\maketitle
\end{CJK}
Along an isotopic chain, with increasing neutron number the nuclei gradually lose binding as the dripline (the limit of nuclear existence) is approached \cite{Otsuka19,Forssen13}. When occupying the $s$ or $p$ orbital, the least bound neutrons of near-dripline nuclei can tunnel far out into the ``classically forbidden" region, and a novel phenomenon---the neutron halo---occurs \cite{Tanihata85, Hansen84, Hansen95,Jensen04, Fred12, Tanihata13, Riisager13, Meng06}. It is a prime example of the emergent simplicity in nuclear many-body systems on top of the complexity of interactions among the constituent nucleons \cite{Anderson72,Freer18,Hagen16,Ruiz20}.

In the context of halo, of particular interest are nuclei with a 2$n$-halo structure, which generally exhibit a Borromean character without any bound binary subsystems \cite{Hansen84, Hansen95, Jensen04, Fred12, Tanihata13, Riisager13}. Recently, $2n$-halo structure was reported in $^{22}$C and $^{29}$F from the large matter radius \cite{Tanaka10,Togano16,Bagchi20}  and in $^{19}$B from the enhanced electric dipole strength measurements \cite{Cook20}. It has been of great interest to search for new 2$n$-halo systems \cite{Tanihata13} and new halo features such as the core-halo shape decoupling \cite{Zhou10, Sun18, Pei13} and the Efimov state \cite{Hammer10, Hagen13}. Meanwhile, it is also important to have comprehensive investigation of the structure of a variety of nuclei with both well- and less-developed halos, particularly using transfer or knockout reactions, to reach a detailed understanding of 2$n$ halo \cite{Jensen04, Fred12, Riisager13, Tanihata13}. Such measurements have so far only been made for light systems with a well-developed 2$n$ halo, $^{6}$He \cite{Aumann99,Wang02,Markenroth01}, $^{11}$Li \cite{Zinser97,Simon99,Nakamura06,Tanihata08,Aksyutina13,Sanetullaev16,Kubota17,Kubota20}, $^{14}$Be \cite{Labiche01,Kondo10,Aksyutina13,Aksyutina13b,Corsi19}. In all these cases, the valence neutrons dominantly occupy $p$ and $s$ orbitals. This naturally raises the question whether the dominant occupation of $s$ or $p$ orbitals is universal for $2n$-halo nuclei and should thus be a criterion to identify 2$n$-halo systems \cite{Jensen04, Fred12}. On the other hand, recent theoretical calculations show that a slight $s$-wave tail should be sufficient for the occurrence of $2n$ halo in very weakly bound neutron-rich systems \cite{Hove20}.  

In this Letter, we report the observation of a surprisingly small $s$-orbital component in the Borromean nucleus $^{17}$B, which has long been considered as a 2$n$-halo system \cite{Tanihata13}. Halo features in $^{17}$B have already been reported---the large matter radius  \cite{Suzuki99, Ozawa01}, the narrow momentum distribution of $^{15}$B \cite{Suzuki02}, and the thick neutron surface \cite{Estrade14}, but the $s$-orbital percentage has hitherto not been directly measured. Using a $^{15}$B+$n$+$n$ 3-body model with $^{15}$B being an inert and spherical core, a large $s$-orbital percentage was deduced from the matter radius and the $^{15}$B momentum distribution---36(19)$\%$ \cite{Suzuki99}, 69(20)$\%$ \cite{Suzuki02}, 50(10)$\%$ \cite{Yamagu04},  53(21)$\%$ \cite{Fortune12}. However, Estrad\'{e} {\it et al.} found its neutron skin thickness did not fit in with such a 3-body picture bearing out a dominant $s$-orbital component \cite{Estrade14}. The large $s$-orbital percentage in the neighboring isotopes---$^{14}$B (64$\%$ $\sim$ 89$\%$) \cite{Guim00,Sauvan04,Bedoor13}, $^{15}$B ($\sim$63$\%$) \cite{Sauvan04}, $^{18}$B \cite{Spyrou10}, and $^{19}$B ($\sim$35$\%$) \cite{Cook20}---also makes the $s$-orbital percentage in $^{17}$B an intriguing question.

In the present work, we have achieved the first direct measurement of the $s$-orbital percentage in $^{17}$B using the quasi-free $(p,pn)$ reaction in inverse kinematics. This study concerns a kinematically complete measurement, which was made possible by combining the high-intensity beams provided by the Radioactive Isotope Beam Factory of  RIKEN Nishina Center and the state-of-the-art detector instruments including the vertex-tracking liquid hydrogen target MINOS \cite{AO14}, in-beam $\gamma$-ray spectrometer DALI2 \cite{Tak14}, and the SAMURAI spectrometer \cite{Koba13,Shimizu13,Kondo20}.

\textit{Experiment.\textemdash} Secondary $^{17}$B beams ($\sim$1.4${\times}$10$^{4}$ $pps$, $\sim$277 MeV/nucleon) were produced from the fragmentation of $^{48}$Ca at 345 MeV/nucleon and prepared using the BigRIPS fragment separator \cite{Kubo03, Kubo12}. They were then tracked onto the 150 mm-thick MINOS target \cite{AO14} using two multi-wire drift chambers. At the target region, we placed a $\gamma$-ray detector array constructed with 68 NaI crystals of DALI2 \cite{Tak14},  a recoil-proton spectrometer composted of a multi-wire drift chamber and a plastic scintillator array,  and  the recoil-neutron detector array WINDS \cite{Yasuda16}. The charged fragments and decay neutrons were detected by SAMURAI \cite{Koba13,Shimizu13} and NEBULA \cite{Kondo20}. The relative energy $E_{\rm rel}$ of the unbound nucleus $^{16}$B was reconstructed from the momenta of $^{15}$B and the decay neutron, with a resolution (FWHM) of $\sim$0.45$\sqrt{E_{\rm rel}}$ (in MeV). When $^{15}$B is in an excited state, the energy of  $^{16}$B ($E_{\rm d}$) with respect to the $^{15}$B$(g.s.) + n$  threshold can be obtained as the sum of  $E_{\rm rel}$ and the excitation energy of $^{15}$B. Population of excited $^{15}$B fragments was reported in a prior breakup experiment of $^{17}$B \cite{Kanumgo05}. Details of the setup can be found in \cite{Corsi19,Kubota20,Kubota17}. 

\begin{figure}[htp]
\includegraphics[width=0.99\linewidth]{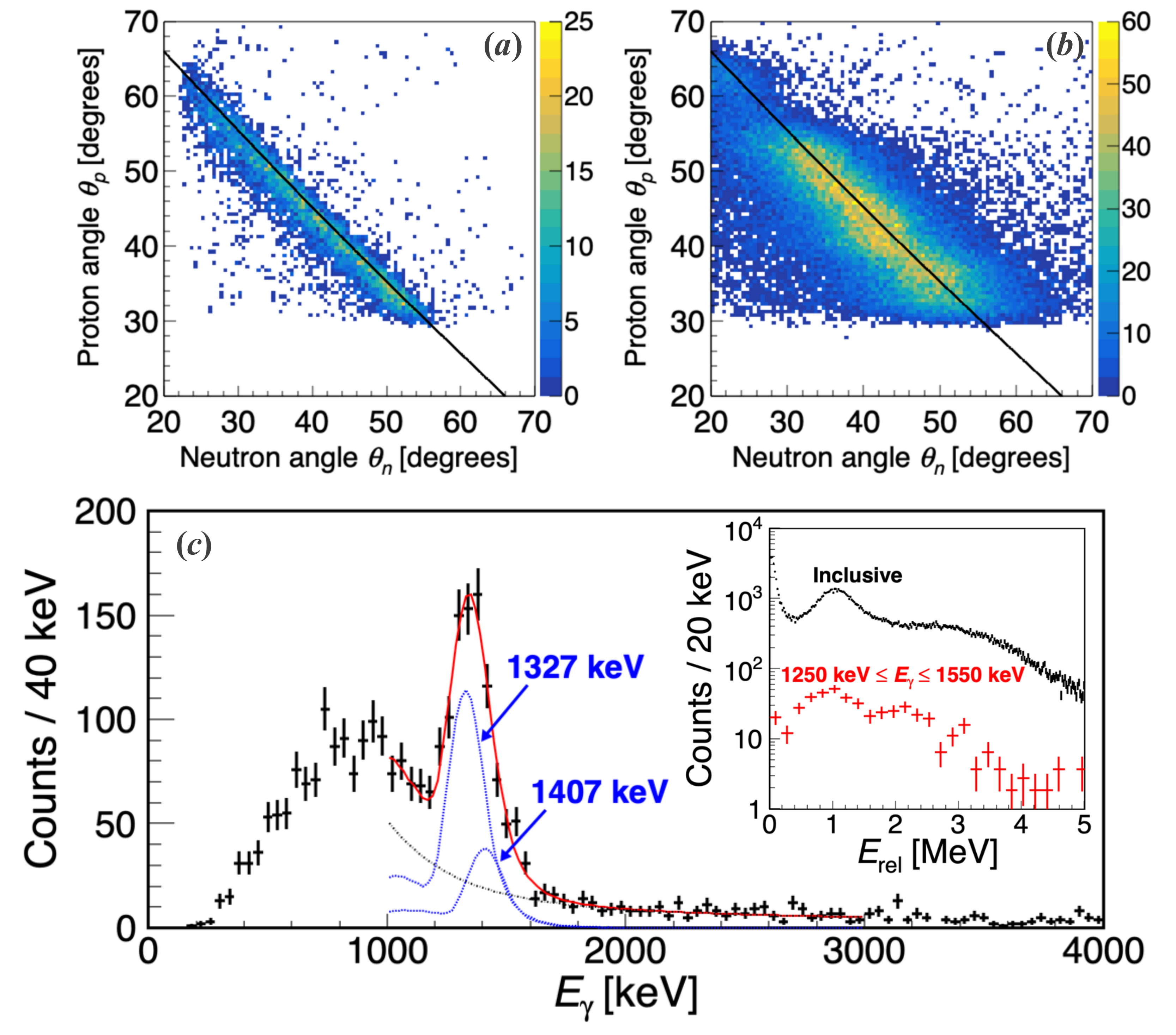}
\caption{Angular correlation between recoil protons and neutrons for events with recoil neutrons detected by WINDS (a) and reconstructed from momentum conservation (b). The black line indicates the kinematical simulation assuming quasi-free $(p,pn)$ off $^{17}$B. (c) Doppler-corrected $\gamma$-ray spectrum, fitted using two $\gamma$ rays (1327 keV and 1407 keV) and a two-exponential background. The inset presents the $\gamma$-gated $E _{\rm rel}$ spectrum, together with the inclusive one for comparison.}
\label{fmech}
\end{figure}

\textit{Quasi-free $(p,pn)$.\textemdash}We first confirmed the quasi-free $(p,pn)$ process by checking the kinematical correlation between recoil protons and recoil neutrons \cite{Corsi19,Panin16,Atar18}. The correlation between the polar angles, $\theta _{p}$ and $\theta _{n}$, agrees nicely with the kinematical simulation (Fig. \ref{fmech}(a)).

The kinematically complete measurement allows us to reconstruct the momentum of the recoil neutron from momentum conservation without detecting it, which largely enhances the statistics. The corresponding angular correlation is presented in Fig. \ref{fmech}(b).  The correlation locus gets broadened relative to panel (a), but follows well the expected correlation pattern of quasi-free $(p,pn)$.  

\textit{Gamma-coincidence.\textemdash}The Doppler-shift-corrected $\gamma$-ray spectrum in coincidence with $^{16}$B ($E _{\rm rel}\leq$ 5 MeV) is shown in Fig. \ref{fmech}(c).  It is well fitted using the response functions of two known $\gamma$ rays, 1327 keV and 1407 keV from $^{15}$B excited states \cite{Stanoiu04,Kanumgo05,Kondo05}, and a two-exponential background ($\chi^{2}/ndf$=1.4).  For each $\gamma$ ray, the response function was obtained from Geant4 simulations considering the realistic setup and the resolution of each crystal. The inset shows the $\gamma$-gated $E _{\rm rel}$ spectrum after correcting for the $\gamma$ efficiency ($\sim$12$\%$), in comparison to the inclusive one. Obviously, the core-excited component is small (within $\sim$5$\%$), and thus neglected when fitting the $E_{\rm rel}$ spectrum (see below).  $^{16}$B states from $\gamma$-coincident analysis  \cite{Suppl} are presented in Table \ref{results}.

\textit{$E_{\rm rel}$ spectrum of $^{16}$}B.\textemdash For $^{17}$B with $N$=12, the knocked-out neutron should mainly come from 1$\emph{s}_{1/2}$ and 0$\emph{d}_{5/2}$ orbitals. The small contribution from $p$-wave orbitals was confirmed by checking the angular correlation \cite{Simon99}.  Ignoring higher-lying 0$\emph{d}_{3/2}$ is justified by the theoretical calculations we employed (see below).

\begin{figure}[thbp]
\includegraphics[width=0.95\linewidth]{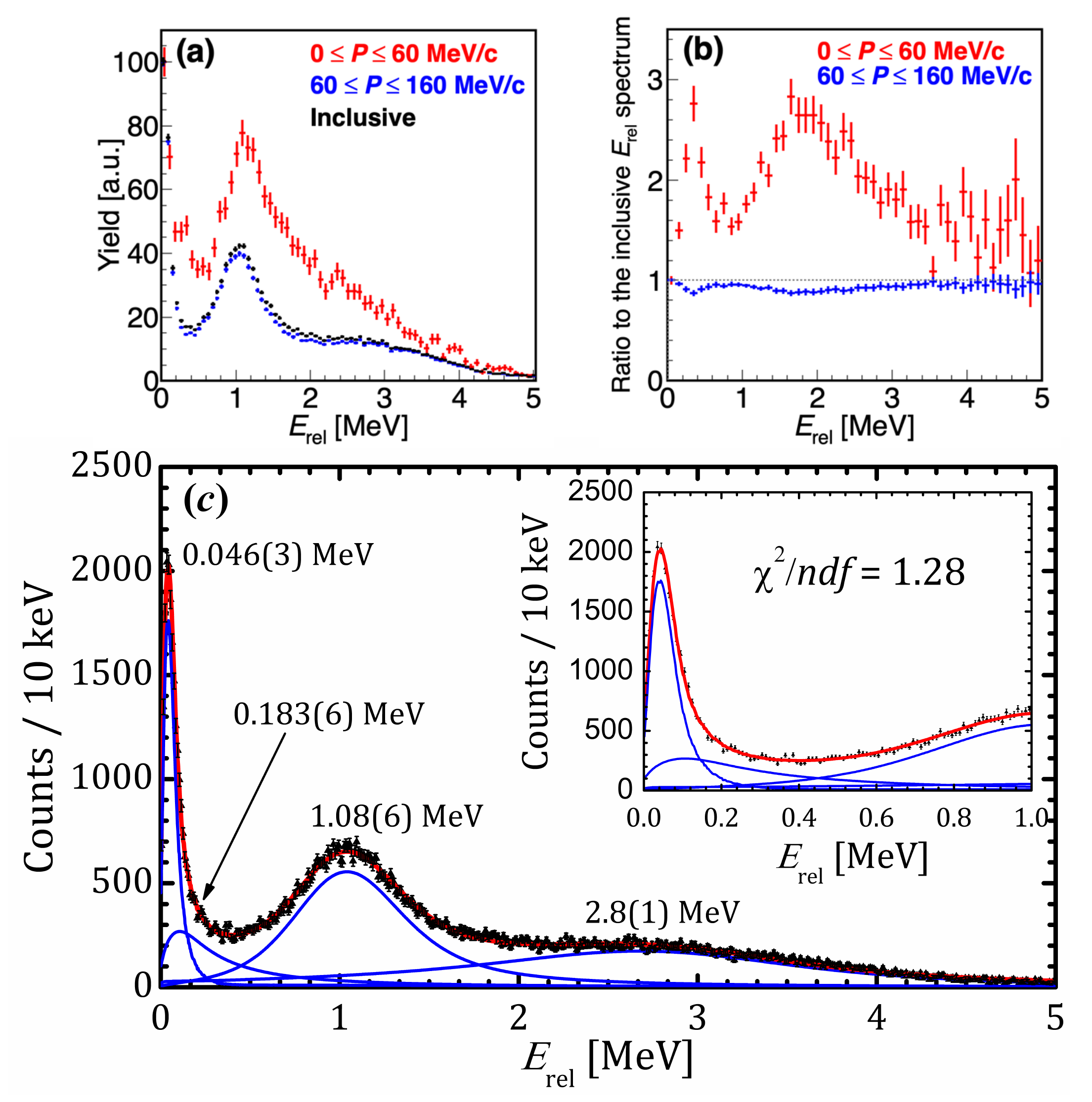}
\caption{(a)$^{16}$B $E_{\rm rel}$ spectra gated by 0 MeV/$c$ $\leq P \leq60$ MeV/$c$ (red) and  60 MeV/$c$ $\leq P \leq160$ MeV/$c$ (blue) in comparison to the inclusive one (black). (b) Ratios of the momentum-gated spectrum to the inclusive one. The grey dashed line, with a constant value of 1, stands for the scenario that the entire spectrum is associated with knockout of 0$\emph{d}_{5/2}$ neutrons. (c) The fitting with a sum of four resonances. The inset is a zoom-in view of the 0-1 MeV region.}
\label{fpana}
\end{figure}

We first checked the $E_{\rm rel}$ spectrum gated by the momentum ($P$) of the knocked-out neutron to disentangle states populated by 1$\emph{s}_{1/2}$ and 0$\emph{d}_{5/2}$ knockout. The $E_{\rm rel}$ spectra gated by 0 MeV/$c$ $\leq P \leq60$ MeV/$c$ (selective for 1$\emph{s}_{1/2}$) and  60 MeV/$c$ $\leq P \leq160$ MeV/$c$ (selective for 0$\emph{d}_{5/2}$) are shown in Fig. \ref{fpana}(a), together with the inclusive one for comparison. All the spectra are normalized according to the peak at $\sim$0, since it is 3$^{-}$ associated purely with 0$\emph{d}_{5/2}$ (see below). Its $d$-wave character has also been established in \cite{Lecou09}. To enhance the visibility,  in Fig. \ref{fpana}(b) we presented the ratio of the $P$-gated spectrum to the inclusive one. Obviously, the prominent peak at $\sim$1 MeV in Fig. \ref{fpana}(a) is mainly from 0$\emph{d}_{5/2}$ knockout. Meanwhile, the red histogram in Fig. \ref{fpana}(b) clearly shows two 1$\emph{s}_{1/2}$-associated states, one at $\sim$0.2 MeV and another broad bump at 1$\sim$4 MeV. 

%---$d$-wave Breit-Wigner line shapes for the  states at $\sim$0.04 MeV and $\sim$1 MeV and $s$-wave line shapes for the other two \cite{Lane58}---
Hence, the $E_{\rm rel}$ spectrum was fitted using four resonances, after taking into account the experimental acceptance and resolutions. The states at $\sim$0.04 MeV and $\sim$1 MeV were described with $d$-wave Breit-Wigner line shapes and the other two with $s$-wave line shapes  \cite{Lane58}. Since the intrinsic width of the $\sim$0.04-MeV state is much smaller than the experimental resolution, it was fixed to 1 keV in the fitting, and an upper limit of $\sim$20 keV was estimated. The results are presented in Fig.~\ref{fpana}(c) and Table \ref{results}. Errors of the resonance parameters include both statistical and systematic errors, the latter being dominated by the effect of the fitting range. The 0.046(3)-MeV state agrees with prior reports \cite{Lecou09, Spyrou10, Kalpa00}, but the resonant energy is refined; meanwhile, the inset of Fig. \ref{fmech}(c) clearly shows that it is associated with $^{15}$B$(g.s.)$. The 2.38(8)-MeV state observed in the $\gamma$-coincident analysis also agrees well with the reported state at 2.40(7) MeV \cite{Kalpa00}.  

\begin{table*}[t]
\small
\caption{\label{results} Summary of $^{16}$B states and experimental spectroscopic factors ($S_{\rm exp}$). The experimental ($\sigma _{\rm exp}$) and theoretical single-particle cross sections ($\sigma _{\rm th}$) are the integrated value over the detector coverage. $S_{\rm exp}$ is defined as $S_{\rm exp}=\sigma _{\rm exp}/\sigma _{\rm th}$. }
\centering
\begin{threeparttable}
\begin{tabular*}{1\linewidth}{@{\extracolsep{\fill}}ccccccccc}
\hline
\hline
 \multirow{2}*{\emph{$E_{\rm x}(^{15}$}B)  [MeV]}&  \multirow{2}*{ \emph{$E_{\rm rel}$}   [MeV]}& \multirow{2}*{ \emph{$\Gamma_{\rm r}$}  [MeV] }& \multirow{2}*{ \emph{$E_{\rm d}$}   [MeV] \tnote{a}}& \multirow{2}*{ $J^{\pi}$ }&  \multirow{2}*{$n$ orbital }&  \multirow{2}*{\emph{$\sigma _{\rm exp}$}  [mb]  }&  \multirow{2}*{\emph{$\sigma _{\rm th}$}  [mb]  }& \multirow{2}*{ \emph{$S_{\rm exp}$} }\\
 &       &        &         &       &      &   & \\
\hline
0 & 0.046(3) & $<$ 0.02&  0.046(3)&$3^{-}$&$0d_{5/2}$& 0.0639(6) & 0.199&0.32(4)\\
\hline
\multirow{2}*{0} & \multirow{2}*{0.183(6)} & \multirow{2}*{0.44(7)}& \multirow{2}*{0.183(6)} &\multirow{2}*{$2^{-}$}& $0d_{5/2}$&0.041(5) &0.196&0.21(3)\\
\cline{6-9}
~ & ~ & ~& ~ &~& $1s_{1/2}$&0.007(2)&0.329&0.02(1)\\
\hline
0  & 1.08(6)& 0.5(2)&1.08(6)  &$4^{-}$&$0d_{5/2}$ &0.18(2)&0.181&0.97(14)\\
\hline
\multirow{2}*{0} & \multirow{2}*{ 2.8(1)} & \multirow{2}*{1.8(3)}& \multirow{2}*{2.8(1)} &\multirow{2}*{$1^{-}$/ $2^{-}$}& $0d_{5/2}$&0.14(2)&0.160&0.87(14)\\
\cline{6-9}
~ & ~ & ~& ~ &~& $1s_{1/2}$&0.054(9)&0.243&0.22(4)\\
\hline
1.33 & 1.05(8) & 0.3(2) & 2.38(8) &$(3^{-})$& $0d_{5/2}$  & 0.009(3)&0.164&0.06(2)\\
\hline
1.33 & 2.4(2) & 0.4(3)&\multirow{2}*{3.7(2)}\tnote{b}&\multirow{2}*{$(3^{-}$/ $4^{-}$) }& \multirow{2}*{$0d_{5/2}$}  & \multirow{2}*{0.011(4)}& \multirow{2}*{0.154}& \multirow{2}*{0.07(3)}\\
\cline{1-3}
2.73 & 0.9(3) & 0.5(2) & ~&~& ~  & ~& ~& ~\\
\hline
2.73 & 2.1(1) &  $<$ 0.5& 4.8(1) &$(3^{-}$/ $4^{-}$)& $0d_{5/2}$  & 0.003(2)&0.144&0.02(1)\\
\hline
\multicolumn{5}{c}{ } & \multicolumn{3}{c}{$1s_{1/2}$ spectroscopic factor percentage} & 9(2) $\%$\\
\hline
\hline
\end{tabular*}
\begin{tablenotes}
\footnotesize
\item[a]{$E_{\rm d}$=$E_{\rm rel}$+$E_{\rm x}$($^{15}$B).}
\item[b]{Taken as the weighted average of the results of the two decay channels.}
\end{tablenotes}
\end{threeparttable}
\end{table*}

\begin{figure}[htp]
\includegraphics[width=0.9\linewidth]{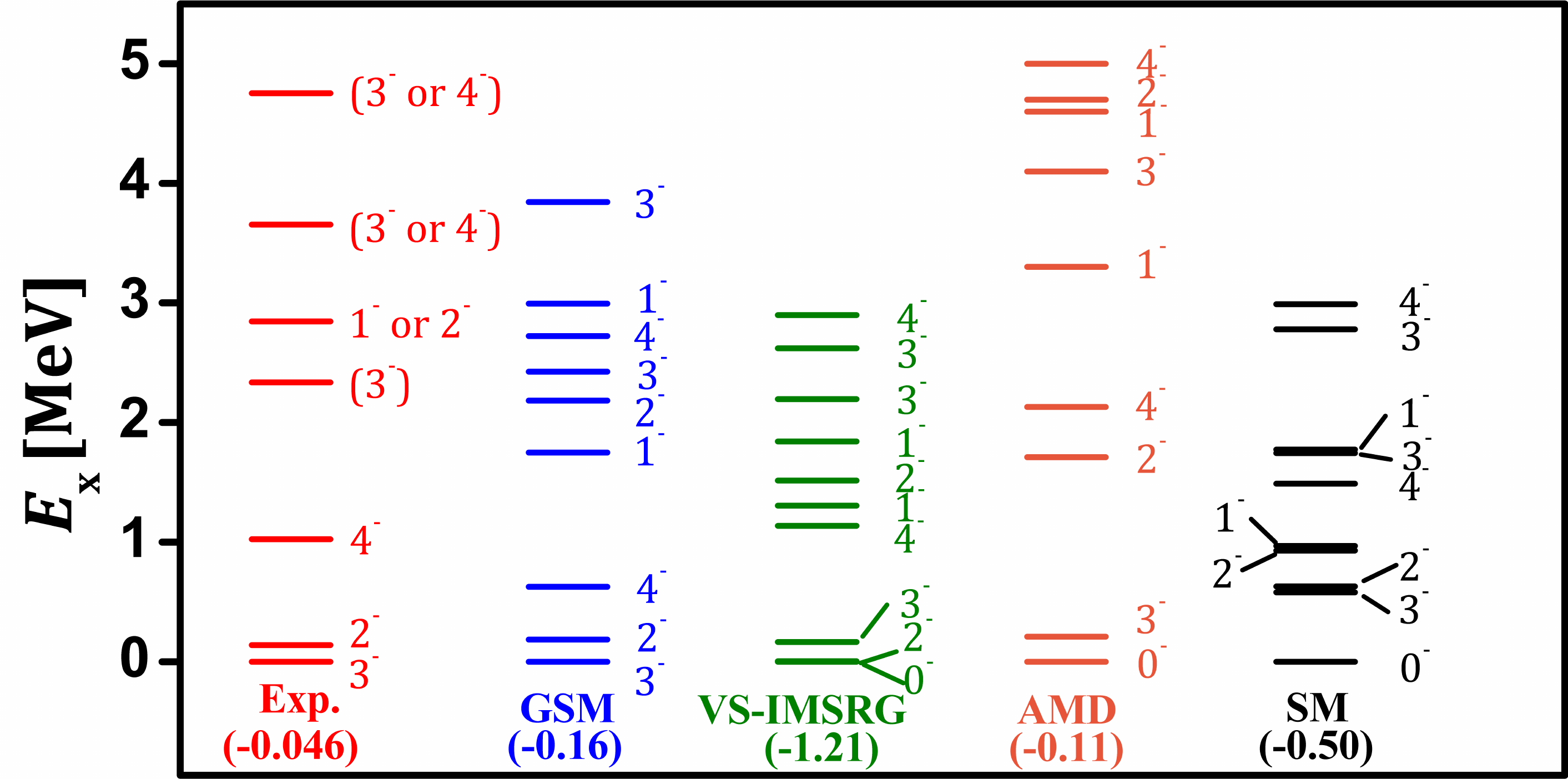}
\caption{Observed $^{16}$B states compared to Gamow shell model (GSM), valence-space in-medium similarity renormalization group (VS-IMSRG), antisymmetrized molecular dynamics (AMD), and shell model (SM) calculations. The $1n$-separation energies are shown in the parentheses (in MeV). }
\label{Theory}
\end{figure}

\textit{Results and Discussions.\textemdash}In Fig.~\ref{Theory}, we compare the level scheme of  $^{16}$B to theoretical calculations: $(i)$ Shell model (SM) calculation with YSOX interactions, considering the reduction of $sd$-shell $n$-$n$ interactions by a factor of 0.75 \cite{Yuan12a,Yuan12b,Sta08,Soh08,Ueno96}.  $(ii)$ Valence-space in-medium similarity renormalization group (VS-IMSRG) calculation \cite{Stroberg13} using the optimized chiral effective field theory interaction at next-to-next-to-leading order \cite{Eks13} in Hartree-Fock basis with 15 major harmonic-oscillator shells ($\hbar\omega$ = 24 MeV). $(iii)$ Antisymmetrized molecular dynamics plus generator coordinate method (AMD) calculation using the Gogny D1S interaction \cite{Estrade14, Kimura04}. $(iv)$ Gamow shell model (GSM) calculation with a $^{10}$He core \cite{Michel09,Michel02,Michel20}.  All partial waves up to $l$ = 3 were included for the valence neutrons, while $\emph{p}_{3/2}$ and $\emph{p}_{1/2}$ basis states were included for the well-bound valence protons. For the two-body force we adopted the Minnesota force \cite{Thompson77}, and the one-body force was modeled with a Woods-Saxon potential with the potential parameters adjusted to reproduce the separation energies and low-lying states of $^{14}$B and $^{15}$B.

Given that the spin-parity ($J^{\pi}$) of $^{17}$B$(g.s)$ is $3/2^{-}$, $J^{\pi}$ of the two $1s _{1/2}$-associated states (0.183 MeV, 2.8 MeV) should be either $1^{-}$ or $2^{-}$, while the other two non-$1s _{1/2}$ states should be more likely $3^{-}$ or $4^{-}$. Tentative assignments of $J^{\pi}$ can thus be made, as shown in Fig. \ref{Theory}. Though predicted as the ground state by SM, AMD, and VS-IMSRG, 0$^{-}$ should be safely excluded for the 0.046 MeV state, because of the small spectroscopic factors ($<$0.03) predicted by these models. 

In order to analyze the momentum distribution of the knocked-out neutron and extract the $0d _{5/2}$ and $1s _{1/2}$ spectroscopic factors in $^{17}$B, we carried out the distorted-wave impulse approximation (DWIA) calculation \cite{Wakasa17}. This model has recently been applied in several $(p,pn)$ and $(p,2p)$ experiments \cite{Kubota17,Kubota20,Taniuchi19,Chen19,Cortes19,Elekes19,Sun20,Lokotko20,Tang20}. The single-particle wave function and the nuclear density were obtained using the Bohr-Mottelson potential \cite{Bohr69}. The optical potentials for the distorted waves in the initial and final channels were constructed with the microscopic folding model \cite{Toyokawa13} using the Melbourne g-matrix interaction \cite{Amos00} with the spin-orbit component disregarded. For the $p$-$n$ interaction, the Franey-Love effective interaction \cite{Franey85} was adopted. DWIA calculations were performed separately for all reaction channels listed in Table \ref{results}, taking into account $E_{\rm d}$ of $^{16}$B.

\begin{figure}[htp]
\includegraphics[width=1\linewidth]{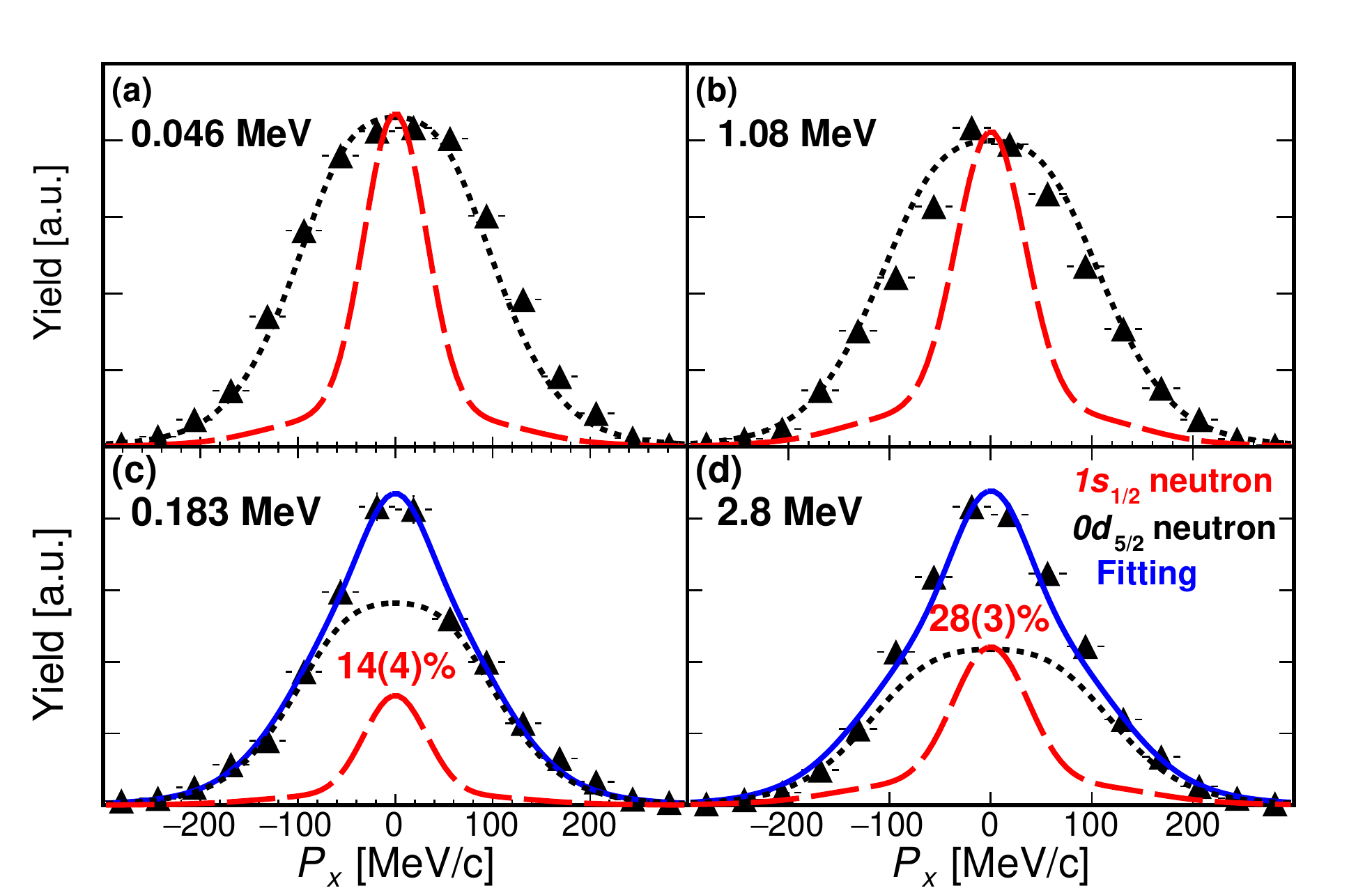}
\caption{Transverse momentum ($P_{\rm x}$) distributions for different $^{16}$B states. The vertical error bars stand for the statistical error, while the horizontal for the bin size.  In (a) and (b), DWIA calculated curves for knockout of $1s _{1/2}$ (red-dashed) and $0d _{5/2}$ (black-dotted) neutrons are normalized to the peak of the experimental spectrum; in (c) and (d), the blue-solid curves represent the fitting with a combination of $1s _{1/2}$ (red-dashed) and $0d _{5/2}$ (black-dotted). All DWIA curves have been convoluted with the experimental resolution.}
\label{momentum}
\end{figure}

Figure \ref{momentum} shows the transverse momentum ($P_{\rm x}$) distributions, together with DWIA calculated curves for $1s _{1/2}$ and $0d _{5/2}$ knockout after folding in the experimental resolution (FWHM) of 45 MeV/$c$. For the 0.046-MeV state (0 MeV $\leq E_{\rm rel} \leq0.1$ MeV) and  the 1.08-MeV state (0.9 MeV $\leq E_{\rm rel} \leq1.3$ MeV), the data can be well reproduced by a pure $0d _{5/2}$ component, in line with the above $J^{\pi}$ assignment of $3^{-}$ and $4^{-}$. 

For the 0.183-MeV state (0.25 MeV $\leq E_{\rm rel} \leq0.45$ MeV) and the 2.8-MeV state (2.5 MeV $\leq E_{\rm rel} \leq3.5$ MeV), we performed the minimum-$\chi^2$ fitting  using a combination of $1s _{1/2}$ and $0d _{5/2}$. A $1s _{1/2}$ fraction of $14(4)\%$ and $28(3)\%$ was obtained, respectively. The errors include both the statistical and systematic errors, the latter being dominated by DWIA and the $E_{\rm rel}$ gate for the data. For the 2.8-MeV state, the same $1s _{1/2}$ fraction is obtained when gating on the left ($29(2)\%$) or right half ($28(2)\%$) of the $E_{\rm rel}$ peak in the analysis, providing further evidence that it is a singlet rather than a doublet. 

For $\gamma$-coincident $^{16}$B states, the $P_{\rm x}$ distributions are all in agreement with knockout of 0$d _{5/2}$ neutrons from $^{17}$B \cite{Suppl}. This naturally leads to a $J^{\pi}$ of  $3^{-}$ or $4^{-}$ , as shown in Fig. \ref{Theory}. The contribution of the $^{14}$B+$2n$ channel is very small ($\sim$5 $\%$) and thus not included in Table \ref{results}, and the $P_{\rm x}$ distribution is consistent with $0d _{5/2}$ knockout.

We then deduced the exclusive cross sections ($\sigma_{\rm exp}$), as tabulated in Table \ref{results}. For the 0.183-MeV and 2.8-MeV states, the $1s _{1/2}$ fraction determined above has been used. The experimental spectroscopic factor ($S_{\rm exp}$) is obtained by dividing $\sigma_{\rm exp}$ with the theoretical cross section for a unit spectroscopic factor ($\sigma_{\rm th}$) from DWIA. Both $\sigma_{\rm exp}$ and $\sigma_{\rm th}$ are the integrated cross sections over the detector coverage ($35^{\circ}<\theta_{p}<55^{\circ}$). We have incorporated the experimental conditions into DWIA to facilitate the direct comparison of $\sigma_{\rm exp}$ and  $\sigma_{\rm th}$ \cite{Suppl}. The errors quoted are the combined statistical and systematic errors. For the 0.046-MeV and 1.08-MeV states, the systematic error of $\sigma_{\rm exp}$ is dominated by the correction of detector efficiencies (9$\%$). For the 0.183-MeV and 2.8-MeV states, the uncertainty of the $1s _{1/2}$ fraction has also been considered. For $S_{\rm exp}$, the uncertainty on $\sigma_{\rm th}$ (within 10$\%$) has further been included, estimated by varying the potential parameters and $E_{\rm d}$ of $^{16}$B  in the DWIA calculation.

The total spectroscopic factors for $1s_{1/2}$  ($S_s$ = 0.24(4)) and $0d_{5/2}$ ($S_d$ = 2.53(21)) are obtained by summing up the $S_{\rm exp}$ for $1s_{1/2}$ and $0d_{5/2}$ respectively in Table \ref{results}, and a $1s_{1/2}$ spectroscopic factor percentage of 9(2)$\%$ is thus deduced from the ratio of $S_s$ and $S_s$ + $S_d$. As shown in Fig.~\ref{Theory}, some $^{16}$B states are not observed in the current quasi-free $(p,pn)$ experiment, indicating that these states should have very small spectroscopic factors in $^{17}$B; their effect on $S_s$ and the $1s_{1/2}$ spectroscopic factor percentage is negligible compared to the errors quoted above. Note that the 1.08-MeV state could be a doublet of two closely located states, given its relatively large width. But such possibility will not affect our conclusion---a small $1s_{1/2}$ component in $^{17}$B, since the $P_{\rm x}$ distribution in Fig. \ref{momentum}(b) clearly shows it is populated (almost) purely by $0d _{5/2}$ knockout. Following the conventional picture for ``$2n$-halo nuclei" \cite{Jensen04}, one may expect a $^{15}$B+2$n$  structure for $^{17}$B \cite{Suzuki99, Suzuki02, Yamagu04, Fortune12}, and the $1s_{1/2}$ spectroscopic factor percentage of 9(2)\% thus leads to a percentage of only 9(2)\% for the halo-relevant $(1s_{1/2})^2$ configuration. As discussed by Estrad\'{e} {\it et al.} \cite{Estrade14}, $^{17}$B may be better described in a $^{13}$B+4$n$ model, and our result would then lead to a percentage of $\sim$18$\%$ for the halo-relevant $(1s_{1/2})^2(0d_{5/2})^2$ configuration, which is also small.

However, prior experiments indeed consistently point to the formation of halo in $^{17}$B \cite{Suzuki99, Ozawa01, Suzuki02, Yamagu04, Estrade14}, which seems to suggest a predominant $s$ orbital component ($\sim$$50\%$) \cite{Suzuki99, Ozawa01, Suzuki02, Yamagu04, Fortune12} in accordance with the conventional picture of ``$2n$-halo nuclei"---an inert core plus two spatially decoupled valence neutrons \cite{Jensen04}. To understand this seeming discrepancy, we compared our result to theoretical predictions obtained by summing up the $1s_{1/2}$ and $0d_{5/2}$  spectroscopic factors of states shown in Fig. \ref{Theory}. The $1s_{1/2}$ percentage is largely overestimated by SM ($25\%$), VS-IMSRG ($26\%$), and GSM ($45\%$). For SM and GSM, we have also checked that the result is not sensitive to the one-body Hamiltonian of $1s_{1/2}$. This may suggest significant impact of deformation or other many-body effects, which can not be sufficiently considered within the shell-model framework. A small $1s_{1/2}$ percentage of $5\%$ is provided by AMD, which is based on the nucleonic degrees of freedom and also predicts a large prolate deformation in $^{17}$B ($\beta$ = 0.4). But the $1s_{1/2}$ percentage is significantly underestimated by AMD, mainly due to the use of Gaussian wave functions \cite{Kimura15}. 

We then resort to the deformed relativistic Hartree-Bogoliubov theory in continuum (DRHBc) \cite{Zhou10,Li12,Li12a, Sun18,Meng15,SunX20}, which self-consistently considers weak binding, deformation, and pairing-induced continuum coupling. We used the effective interaction PK1 \cite{Long04} and a density-dependent zero-range pairing force \cite{Sun18}. The neutron and matter radius \cite{Estrade14}, deformation \cite{Kondo05,Domb05}, and $S_{2n}$ \cite{Wang17} of $^{17}$B  are well reproduced \cite{Suppl}. DRHBc provides a small $1s_{1/2}$ orbital percentage of 14$\%$ for the valence neutrons, close to the experimental result of 9(2)$\%$. Meanwhile, the neutron density distribution shows a slight but definite low-density tail extending into large radial distances \cite{Suppl}, indicating a weak halo component in $^{17}$B. Hence, the neutron halo exists in $^{17}$B as reported in prior experiments \cite{Suzuki99, Ozawa01, Suzuki02, Yamagu04, Estrade14}, but not as the dominant structure component.

\textit{Summary.\textemdash}We have measured the $0d _{5/2}$ and $1s _{1/2}$ spectroscopic factors in $^{17}$B using the quasi-free $(p,pn)$ reaction in inverse kinematics. A small spectroscopic factor percentage of 9(2)$\%$ was determined for the $1s_{1/2}$ orbital. Our result thus reveals a surprisingly small $s$-orbital component in $^{17}$B whether it is described in a simple $^{15}$B+2$n$ model or more properly in a $^{13}$B+4$n$ model. Our finding of such a small $1s_{1/2}$ component and the halo features in $^{17}$B reported in prior experiments \cite{Suzuki99, Ozawa01, Suzuki02, Yamagu04,Estrade14} can be well explained by DRHBc, revealing a  definite but not dominant halo component in $^{17}$B. The present work gives the smallest $s$- or $p$-orbital component among known nuclei exhibiting halo features, and implies that the dominant occupation of $s$ or $p$ orbitals is not a prerequisite for the occurrence of neutron halo. In weakly bound neutron-rich nuclei, as long as $s$ or $p$  orbitals around the Fermi surface are occupied by the least bound neutrons with an appreciable strength, the halo naturally occurs and coexists with other non-halo configurations \cite{Hansen84, Jensen04, Fred12, Tanihata13, Riisager13}. The halo component, whether or not dominant, results in a distinctive diffused surface, and thus manifests itself in reactions sensitive to the surface properties \cite{Jensen04, Fred12, Tanihata13, Riisager13}.

\acknowledgments {We would like to thank the RIBF accelerator staff for the primary beam delivery and the BigRIPS team for their efforts in preparing the secondary beams. Z. H. Y. acknowledges fruitful discussions with C.L. Bai, J. Casal, Y. Sato, T.T. Sun,  and S.M. Wang, and the support from the Foreign Postdoctoral Researcher program of RIKEN. This work has been supported by the European Research Council through the ERC Grant No. MINOS-25856; the JSPS KAKENHI Grants No. JP16K05352, No. JP18H05404, and No. JP16H02179; the National Natural Science Foundation of China under Grants No. 11525524, No. 11835001, No. 11921006, No. 11975282, No. 11435014, No. 11775316, and No. 11961141004; the Strategic Priority Research Program of Chinese Academy of Sciences, Grant No. XDB34000000; the Institute for Basic Science (IBS-R031-D1). The computation of this work was partly supported by the HPC Cluster of ITP-CAS and the Supercomputing Center, CNIC of CAS.}


\begin{thebibliography}{121}
\bibitem{Otsuka19} T. Otsuka, A. Gade, O. Sorlin, T. Suzuki, and Y. Utsuno, Rev. Mod. Phys. \textbf{92}, 015002 (2020).
\bibitem{Forssen13} C. Forssen, G. Hagen, M. Hjorth-Jensen, W. Nazarewicz, and J. Rotureau, Phys. Scr. \textbf{T152}, 014022 (2013).
\bibitem{Tanihata85} I. Tanihata, H. Hamagaki, O. Hashimoto, Y. Shida, N. Yoshikawa, K. Sugimoto, O. Yamakawa, T. Kobayashi, and N. Takahashi,  Phys. Rev. Lett. \textbf{55}, 2676 (1985).
\bibitem{Hansen84} P. G. Hansen and B. Jonson, Europhys. Lett. \textbf{4}, 409 (1987).
\bibitem{Hansen95} P. G. Hansen and A. S. Jensen,  Ann. Rev. Nucl. Part. Sci. \textbf{45}, 591(1995).
\bibitem{Jensen04} A. S. Jensen, K. Riisager, D. V. Fedorov, and E. Garrido,  Rev. Mod. Phys. \textbf{76}, 215(2004).
\bibitem{Fred12} T. Frederico, A. Delfino, L. Tomio, and M. T. Yamashita,  Prog. Part. Nucl. Phys. \textbf{67}, 939 (2012).
\bibitem{Tanihata13} I. Tanihata, H. Savajols, and R. Kanungo,  Prog. Part. Nucl. Phys. \textbf{68}, 215 (2013).
\bibitem{Riisager13} K. Riisager, Phys. Scr. \textbf{T152}, 014001 (2013).
\bibitem{Meng06} J. Meng, H. Toki, S. G. Zhou, S. Q. Zhang, W. Long, and L. S. Geng, Prog. Part. Nucl. Phys. \textbf{57}, 470  (2006).
\bibitem{Anderson72} P. W. Anderson, Science \textbf{177}, 393 (1972).
\bibitem{Hagen16} G. Hagen, M. Hjorth-Jensen, G. R. Jansen, and T. Papenbrock, Phys. Scr. \textbf{91}, 063006 (2016).
\bibitem{Freer18} M. Freer, H. Horiuchi, Y. Kanada-En'yo, D. Lee, and Ulf-G. Mei{\ss}ner, Rev. Mod. Phys. \textbf{90}, 035004 (2018).
\bibitem{Ruiz20} R. F. G. Ruiz and A. R. Vernon, Eur. Phys. J. A  \textbf{56}, 136 (2020).
\bibitem{Tanaka10} K. Tanaka,  T. Yamaguchi, T. Suzuki, T. Ohtsubo, M. Fukuda, D. Nishimura, M. Takechi, K. Ogata, A. Ozawa, T. Izumikawa {\it et al.}, Phys. Rev. Lett. \textbf{104}, 062701 (2010).
\bibitem{Togano16} Y. Togano, T. Nakamura, Y. Kondo, J. A. Tostevin, A. T. Saito, J. Gibelin, N. A. Orr, N.L. Achouri, T. Aumann, H. Baba {\it et al.}, Phys. Lett. B \textbf{761}, 412 (2016).
\bibitem{Bagchi20} S. Bagchi,  R. Kanungo, Y. K. Tanaka, H. Geissel, P. Doornenbal, W. Horiuchi, G. Hagen, T. Suzuki, N. Tsunoda, D. S. Ahn {\it et al.}, Phys. Rev. Lett. \textbf{124}, 222504 (2020).
\bibitem{Cook20} K. J. Cook,  T. Nakamura, Y. Kondo, K. Hagino, K. Ogata, A. T. Saito, N. L. Achouri, T. Aumann, H. Baba, F. Delaunay {\it et al.}, Phys. Rev. Lett. \textbf{124}, 212503 (2020).
\bibitem{Zhou10} S.-G. Zhou, J. Meng, P. Ring, and E.-G. Zhao, Phys. Rev. C \textbf{82}, 011301(R) (2010);
\bibitem{Sun18} X.-X. Sun, J. Zhao, and S.-G. Zhou, Phys. Lett. B \textbf{785}, 530 (2018).
\bibitem{Pei13} J. C. Pei, Y. N. Zhang, and F. R. Xu, Phys. Rev. C \textbf{87}, 051302(R)(2013).
\bibitem{Hammer10} H.-W. Hammer, and L. Platter, Annu. Rev. Nucl. Part. Sci. \textbf{60}, 207(2010).
\bibitem{Hagen13} G. Hagen, P. Hagen, H.-W. Hammer, and L. Platter, Phys. Rev. Lett. \textbf{111}, 132501 (2013).
\bibitem{Aumann99} T. Aumann, D. Aleksandrov, L. Axelsson, T. Baumann, M. J. G. Borge, L. V. Chulkov, J. Cub, W. Dostal, B. Eberlein, T.W. Elze, H. Emling {\it et al.}, Phys. Rev. C \textbf{59}, 1252 (1999).
\bibitem{Wang02} J. Wang, A. Galonsky, J. J. Kruse, E. Tryggestad, R. H. White-Stevens, P. D. Zecher, Y. Iwata, K. Ieki, A. Horv\'{a}th, F. De\'{a}k {\it et al.}, Phys. Rev. C \textbf{65}, 034306 (2002).
\bibitem{Markenroth01} K. Markenroth,  M. Meister, B. Eberlein, D. Aleksandrov, T. Aumann, L. Axelsson, T. Baumann, M.J.G. Borge, L.V. Chulkov, W. Dostal {\it et al.}, Nucl. Phys. A \textbf{679}, 462 (2001).
\bibitem{Zinser97} M. Zinser, E Humbert, T. Nilsson, W. Schwab, H. Simon, T. Aumann, M.J.G. Borge, L.V. Chulkov, J. Cub, Th. W. Elze {\it et al.}, Nucl. Phys. A \textbf{619}, 151 (1997).
\bibitem{Simon99} H. Simon, D. Aleksandrov, T. Aumann, L. Axelsson, T. Baumann, M. J. G. Borge, L. V. Chulkov, R. Collatz, J. Cub, W. Dostal {\it et al.}, Phys. Rev. Lett.  \textbf{83}, 496 (1999).
\bibitem{Nakamura06} T. Nakamura, A. M. Vinodkumar, T. Sugimoto, N. Aoi, H. Baba, D. Bazin, N. Fukuda, T. Gomi, H. Hasegawa, N. Imai {\it et al.}, Phys. Rev. Lett. \textbf{96}, 252502 (2006).
\bibitem{Tanihata08} I. Tanihata, M. Alcorta, D. Bandyopadhyay, R. Bieri, L. Buchmann, B. Davids, N. Galinski, D. Howell, W. Mills, S. Mythili {\it et al.},  Phys. Rev. Lett. \textbf{100}, 192502 (2008).
\bibitem{Sanetullaev16} A. Sanetullaev, {\it et al.}, Phys. Lett. B \textbf{755}, 481 (2016).
\bibitem{Kubota17} Y. Kubota,  PhD thesis, University of Tokyo, 2017.
\bibitem{Kubota20} Y. Kubota, A. Corsi, G. Authelet, H. Baba, C. Caesar, D. Calvet, A. Delbart, M. Dozono, J. Feng, F. Flavigny {\it et al.},  Phys. Rev. Lett. \textbf{125}, 252501 (2020).
\bibitem{Aksyutina13} Y. Aksyutina, T. Aumann, K. Boretzky, M. J. G. Borge, C. Caesar, A. Chatillon, L. V. Chulkov, D. CortinaGil, U. D. Pramanik, H. Emling {\it et al.}, Phys. Lett. B \textbf{718}, 1309 (2013).
\bibitem{Labiche01} M. Labiche, N. A. Orr, F. M. Marqu\'{e}s, J. C. Ang\'{e}lique, L. Axelsson, B. Benoit, U. C. Bergmann, M. J. G. Borge, W. N. Catford, S. P. G. Chappell {\it et al.},  Phys. Rev. Lett. \textbf{86}, 600 (2001).
\bibitem{Kondo10} Y. Kondo, T. Nakamura, Y. Satou, T. Matsumoto, N. Aoi, N. Endo, N. Fukuda, T. Gomi, Y. Hashimoto, M. Ishihara {\it et al.},  Phys. Lett. B  \textbf{690}, 245 (2010).
\bibitem{Aksyutina13b} Y. Aksyutina, T. Aumann, K. Boretzky, M. J. G. Borge, C. Caesar, A. Chatillon, L. V. Chulkov, D. Cortina-Gil, U. Datta Pramanik, H. Emling {\it et al.}, Phys. Rev. C \textbf{87}, 064316 (2013).
\bibitem{Corsi19} A. Corsi, Y. Kubota, J.Casal, M. G\'{o}mez-Ramos, A. M. Moro, G. Authelet, H.Baba, C.Caesar, D.Calvet, A.Delbart {\it et al.}, Phys. Lett. B \textbf{797}, 134843 (2019).
\bibitem{Hove20} D. Hove,E. Garrido, P. Sarriguren, D. V. Fedorov, H. O. U. Fynbo, A. S. Jensen, and N. T. Zinner, Phys. Rev. Lett. \textbf{120}, 052502 (2018).

\bibitem{Ozawa01} A. Ozawa, T. Suzuki, I. Tanihata, Nucl. Phys. A \textbf{693}, 32 (2001).
\bibitem{Suzuki99} T. Suzuki, T. Suzuki, R. Kanungo, O. Bochkarev, L. Chulkov, D. Cortina, M. Fukuda, H. Geissel, M. Hellstrom, M. Ivanov, R. Janik {\it et al.}, Nucl. Phys. A \textbf{658}, 313 (1999).
\bibitem{Suzuki02} T. Suzuki, Y. Ogawa, M. Chiba, M. Fukuda, N. Iwasa, T. Izumikawa, R. Kanungo, Y. Kawamura, A. Ozawa, T. Suda {\it et al.}, Phys. Rev. Lett. \textbf{89}, 012501 (2002).
\bibitem{Estrade14} A. Estrad\'{e}, R. Kanungo, W. Horiuchi, F. Ameil, J. Atkinson, Y. Ayyad, D. Cortina-Gil, I. Dillmann, A. Evdokimov, F. Farinon {\it et al.}, Phys. Rev. Lett. \textbf{113}, 132501 (2014).
\bibitem{Yamagu04} Y. Yamaguchi, C. Wu,T. Suzuki, A. Ozawa, D. Q. Fang, M. Fukuda, N. Iwasa, T. Izumikawa, H. Jeppesen, R. Kanungo {\it et al.}, Phys. Rev. C \textbf{70}, 054320 (2004).
\bibitem{Fortune12} H. T. Fortune and R. Sherr, Eur. Phys. J. A \textbf{48}, 103 (2012). 
\bibitem{Guim00} V. Guimar\~{a}es, J. J. Kolata, D. Bazin, B. Blank, B. A. Brown, T. Glasmacher, P. G. Hansen, R. W. Ibbotson, D. Karnes, V. Maddalena {\it et al.}, Phys. Rev. C \textbf{61}, 064609 (2000).
\bibitem{Sauvan04} E. Sauvan, F. Carstoiu, N. A. Orr, J. S. Winfield, M. Freer, J. C. Ang\'{e}lique, W. N. Catford, N. M. Clarke, M. MacCormick, N. Curtis {\it et al.}, Phys. Rev. C \textbf{69}, 044603 (2004).
\bibitem{Bedoor13} S. Bedoor, A. H. Wuosmaa, J. C. Lighthall, M. Alcorta, B. B. Back, P. F. Bertone, B. A. Brown, C. M. Deibel, C. R. Hoffman, S. T. Marley {\it et al.}, Phys. Rev. C \textbf{88}, 011304(R) (2013).
\bibitem{Spyrou10} A. Spyrou, T. Baumann, D. Bazin, G. Blanchon, A. Bonaccorso, E. Breitbach, J. Brown, G. Christian, A. Deline, P. A. DeYoung {\it et al.}, Phys. Lett. B \textbf{683}, 129 (2010).
\bibitem{AO14} A. Obertelli, A. Delbart, S. Anvar, L. Audirac, G. Authelet, H. Baba, B. Bruyneel, D. Calvet, F. Ch\^{a}teau, A. Corsi {\it et al.}  Eur. Phys. J. A  \textbf{50}, 8 (2014).
\bibitem{Tak14} S. Takeuchi, T. Motobayashi, Y. Togano, M. Matsushita, N. Aoi, K. Demichi, H. Hasegawa, and H. Murakami, Nucl. Instrum. Methods ~Phys. Res., Sect. A  \textbf{763} , 596 (2014).
\bibitem{Koba13} T. Kobayashi, N. Chiga, T. Isobe, Y. Kondo, T. Kubo, K. Kusaka, T. Motobayashi, T. Nakamura, J. Ohnishi, H. Okuno {\it et al.},  Nucl. Instrum. Methods Phys. Res., Sect. B  \textbf{317}, 294 (2013).
\bibitem{Shimizu13} Y. Shimizu, H. Otsu, T. Kobayashi, T. Kubo, T. Motobayashi, H. Sato, and K. Yoneda, Nucl. Instrum. Methods Phys. Res. Sect. B 317, 739 (2013).
\bibitem{Kondo20} Y. Kondo, T. Tomai, and T. Nakamura, Nucl. Instrum. Methods Phys. Res., Sect. B  \textbf{463}, 173 (2020).
\bibitem{Kubo03} T. Kubo, Nucl. Instrum. Methods Phys. Res., Sect. B  \textbf{204},  97 (2003).
\bibitem{Kubo12} T. Kubo, D. Kameda, H. Suzuki, N. Fukuda, H. Takeda, Y. Yanagisawa, M. Ohtake, K. Kusaka, K. Yoshida, N. Inabe {\it et al.}, Prog. Theor. Exp. Phys.  \textbf{2012}, 03C003 (2012).
\bibitem{Yasuda16} J. Yasuda, M. Sasano, R. G. T. Zegers, H. Baba, W. Chao, M. Dozono, N. Fukuda, N. Inabe, T. Isobe, G. Jhang {\it et al.},  Nucl. Instrum. Methods Phys. Res., Sect. B  \textbf{376}, 393 (2016).
\bibitem{Kanumgo05} R. Kanungo, Z. Elekes, H. Baba, Z. Dombr\'{a}di, Z. F\"{u}l\"{o}p, J. Gibelin, \'{A}. Horv\'{a}th, Y. Ichikawa, E. Ideguchi, N. Iwasa {\it et al.}, Phys. Lett. B  \textbf{608}, 206 (2005).
\bibitem{Atar18} L. Atar, S. Paschalis, C. Barbieri, C. A. Bertulani, P. D\'{i}az Fern\'{a}ndez, M. Holl, M. A. Najafi, V. Panin, H. Alvarez-Pol, T. Aumann {\it et al.}, Phys. Rev. Lett. \textbf{120}, 052501 (2018).
\bibitem{Panin16} V. Panin, J. T. Taylor, S. Paschalis, F. Wamers, Y. Aksyutina, H. Alvarez-Pol, T. Aumann, C. A. Bertulani, K. Boretzky, C.Caesar, {\it et al.}, Phys. Lett. B  \textbf{753}, 204 (2016).
\bibitem{Stanoiu04} M.~Stanoiu, M. Belleguic, Zs. Dombr\'{a}di, D. Sohler, F.  Azaiez1, B. A. Brown, M. J. Lopez-Jimenez, M. G. Saint-Laurent, O. Sorlin, Yu.-E. Penionzhkevich {\it et al.},  Eur. Phys. J. A  \textbf{22}, 5 (2004).
\bibitem{Kondo05} Y. Kondo, T. Nakamura, N. Aoi, H. Baba, D. Bazin, N. Fukuda, T. Gomi, H. Hasegawa, N. Imai, M. Ishihara {\it et al.}, Phys. Rev. C \textbf{71}, 044611 (2005).
\bibitem{Suppl} See the Supplemental Material for more details of the $\gamma$-coincident analysis, the distorted-wave impulse approximation (DWIA) calculation, and the deformed relativistic Hartree-Bogoliubov theory in continuum (DRHBc) calculation.
\bibitem{Lecou09} J.-L. Lecouey, N. A. Orr, F. M. Marqu\'{e}s, N. L. Achouri, J. -C. Ang\'{e}lique, B. A. Brown, F. Carstoiu, W. N. Catford, N. M. Clarkee, M. Freer {\it et al.}, Phys. Lett. B \textbf{672}, 6 (2009).
\bibitem{Lane58} A. M. Lane and R. G. Thomas, Rev. Mod. Phys. \textbf{30}, 257 (1958).
\bibitem{Kalpa00} R. Kalpakchieva, H.G. Bohlen, W. von Oertzen, B. Gebauer, M. von Lucke-Petsch, T.N. Massey, A.N. Ostrowski, Th. Stolla, M. Wilpert, and Th. Wilpert, Eur. Phys. J. ~A \textbf{7}, 451 (2000).
\bibitem{Yuan12a} C. X. Yuan, T Suzuki, T. Otsuka, F. R. Xu, and N. Tsunoda, Phys. Rev. C \textbf{85}, 064324 (2012).
\bibitem{Yuan12b} C. X. Yuan, C. Qi, and F. R. Xu,  Nucl. Phys. A \textbf{883}, 25 (2012).
\bibitem{Sta08} M.  Stanoiu, D. Sohler, O. Sorlin, F. Azaiez, Zs. Dombr\'{a}di, B. A. Brown, M. Belleguic, C. Borcea, C. Bourgeois, Z. Dlouhy {\it et al.}, Phys. Rev. C \textbf{78}, 034315 (2008).
\bibitem{Soh08} D. Sohler, M. Stanoiu, Zs. Dombr\'{a}di, F. Azaiez, B. A. Brown, M. G. Saint-Laurent, O. Sorlin, Yu.-E. Penionzhkevich, N. L. Achouri, J. C. Ang\'{e}lique {\it et al.}, Phys. Rev. C \textbf{77}, 044303 (2008).
\bibitem{Ueno96} H. Ueno, K. Asahi, H. Izumi, K. Nagata, H. Ogawa, A. Yoshimi, H. Sato, M. Adachi, Y. Hori, and K. Mochinaga {\it et al.}, Phys. Rev. C \textbf{53}, 2142 (1996).
\bibitem{Stroberg13}S. R. Stroberg,  A. Calci, H. Hergert, J. D. Holt, S.K. Bogner, R. Roth, and A. Schwenk, Phys. Rev. Lett. \textbf{118}, 032502 (2017).
\bibitem{Eks13} A. Ekstr\"{o}m,  G. Baardsen, C. Forss\'{e}n, G. Hagen, M. Hjorth-Jensen, G. R. Jansen, R. Machleidt, W. Nazarewicz, T. Papenbrock, J. Sarich, and S. M. Wild, Phys. Rev. Lett. \textbf{110}, 192502 (2013).
\bibitem{Kimura04} M. Kimura, Phys. Rev. C \textbf{69}, 044319 (2004).
\bibitem{Michel09} N. Michel, W. Nazarewicz, M. Ploszajczak, and T. Vertse, J. Phys. G: Nucl. Part. Phys. \textbf{36}, 013101 (2009).
\bibitem{Michel02} N. Michel, W. Nazarewicz, M. Ploszajczak, and K. Bennaceur, Phys. Rev. Lett. \textbf{89}, 042502 (2002).
\bibitem{Michel20} N. Michel, J. G. Li, F. R. Xu, and W. Zuo, Phys. Rev. C \textbf{101}, 031301(R) (2020).
\bibitem{Thompson77} D. Thompson, M. Lemere, and Y. Tang,  Nucl. Phys. A  \textbf{286}, 53 (1977).
\bibitem{Wakasa17} T. Wakasa, K. Ogata, and T. Noro, Prog. Part. Nucl. Phys. \textbf{96}, 32 (2017).
\bibitem{Taniuchi19} R. Taniuchi, C. Santamaria, P. Doornenbal, A. Obertelli, K. Yoneda, G. Authelet, H. Baba, D. Calvet, F. Ch\^{a}teau, A. Corsi {\it et al.}, Nature \textbf{569}, 53 (2019).
\bibitem{Chen19} S. D. Chen, J. Lee, P. Doornenbal, A. Obertelli, C. Barbieri, Y. Chazono, P. Navr\'{a}til, K. Ogata, T. Otsuka, F. Raimondi {\it et al.}, Phys. Rev. Lett. \textbf{123}, 142501 (2019).
\bibitem{Cortes19} M. L. Cort\'{e}s, W. Rodriguez, P. Doornenbal, A. Obertelli, J. D. Holt, S. M. Lenzi, J. Men'{e}ndez, F. Nowacki, K. Ogata, A. Poves {\it et al.}, Phys. Lett. B \textbf{800}, 135071 (2019).
\bibitem{Elekes19} Z. Elekes, \'{A}. Kripk\'{o}, D. Sohler, K. Sieja, K. Ogata, K. Yoshida, P. Doornenbal, A. Obertelli, G. Authelet, H. Baba {\it et al.}, Phys. Rev. C \textbf{99}, 014312 (2019).
\bibitem{Sun20} Y. L. Sun, A.Obertelli, P. Doornenbal, C. Barbieri, Y. Chazono, T. Duguet, H. N. Liu, P. Navr'{a}til, F. Nowacki, K. Ogata {\it et al.}, Phys. Lett. B \textbf{802}, 135215 (2020).
\bibitem{Lokotko20} T. Lokotko, S. Leblond, J. Lee, P. Doornenbal, A. Obertelli, A. Poves, F. Nowacki, K. Ogata, K. Yoshida, G. Authelet {\it et al.}, Phys. Rev. C \textbf{101}, 034314 (2020).
\bibitem{Tang20} T. L. Tang, T. Uesaka, S. Kawase, D. Beaumel, M. Dozono, T. Fujii, N. Fukuda, T. Fukunaga, A. Galindo-Uribarri, S. H. Hwang {\it et al.}, Phys. Rev. Lett. \textbf{124}, 212502 (2020).
\bibitem{Bohr69} A. Bohr and B. Mottelson, Nuclear Structure (Benjamin,New York, 1969), Vol. I.
\bibitem{Toyokawa13} M. Toyokawa, K. Minomo, and M. Yahiro, Phys. Rev. C \textbf{88}, 054602 (2013).
\bibitem{Amos00} K. Amos, P. Dortmans, H. von Geramb, S. Karataglidis, and J. Raynal, Adv. Nucl. Phys. \textbf{25}, 275 (2000).
\bibitem{Franey85} M. A. Franey and W. G. Love, Phys. Rev. C \textbf{31}, 488 (1985).
\bibitem{Kimura15} M. Kimura, T. Suhara, and Y. Kanada-En'yo, Eur. Phys. J. A \textbf{52}, 373 (2016).
\bibitem{Li12a} L. Li, J. Meng, P. Ring, E.-G. Zhao, and S.-G. Zhou, Chin. Phys. Lett.  \textbf{29}, 042101 (2012).
\bibitem{Li12} L. Li, J. Meng, P. Ring, E.-G. Zhao, and S.-G. Zhou, Phys. Rev. C \textbf{85}, 024312 (2012).
\bibitem{Meng15} J. Meng and  S.-G. Zhou, J. Phys. G: Nucl. Part. Phys. \textbf{42}, 093101 (2015).
\bibitem{SunX20} X.-X. Sun, J. Zhao, and  S.-G. Zhou, Nucl. Phys. A \textbf{1003}, 122011 (2020).
\bibitem{Long04} W. Long, J. Meng, N. Van Giai, and S.-G. Zhou, Phys. Rev. C \textbf{69}, 034319 (2004).
\bibitem{Domb05} Zs. Dombr\'{a}di, Z. Elekes, R. Kanungo, H. Baba, Z. F\"{u}l\"{o}p, J. Gibelin, \'{A}. Horv\'{a}th,  E. Ideguchi, Y. Ichikawa, N. Iwasa {\it et al.}, Phys. Lett. B \textbf{621}, 81(2005).
\bibitem{Wang17} M. Wang, G. Audi, F. G. Kondev, W.J. Huang, S. Naimi and X. Xu, Chin. Phys. C \textbf{41}, 030003 (2017).

\end{thebibliography}
\end{document}